\documentstyle[preprint,prl,aps]{revtex} 

\begin{document}
\draft
\title{\begin{flushright}
{\small\hfill CERN-TH/97-245\\
\hfill hep-th/9709101}\\
\end{flushright}
Two-Dimensional QCD in the Wu-Mandelstam-Leibbrandt
Prescription }
\author{Matthias Staudacher \footnote{
matthias@nxth04.cern.ch; Address after September 1, 1997:
Max-Planck Institut f\"{u}r
Gravitationsforschung, Schlaatzweg 1, D-14473 Potsdam, Germany.}
}
\address{
CERN, Theory Division, CH-1211 Geneva 23, Switzerland}
\author{Werner Krauth \footnote{krauth@physique.ens.fr }}
\address{CNRS-Laboratoire de Physique Statistique,
Ecole Normale Sup\'{e}rieure,
24, rue Lhomond,\\ 75231 Paris Cedex 05, France}
\date{Received \today}
\maketitle
\begin{abstract}
We find the exact non-perturbative expression for a simple Wilson loop of
arbitrary shape for $U(N)$ and $SU(N)$ Euclidean or Minkowskian
two-dimensional Yang-Mills
theory (YM$_2$) regulated by the Wu-Mandelstam-Leibbrandt gauge prescription.
The result differs from the standard pure
exponential area-law of YM$_2$, but still exhibits confinement as well
as invariance under area-preserving diffeomorphisms and
generalized axial gauge transformations. We show
that the large $N$ limit is {\it not} a good approximation
to the model at finite $N$ and conclude that Wu's $N=\infty$
Bethe-Salpeter equation for QCD$_2$ should have no
bound state solutions.
The main significance of our results derives from the importance of the
Wu-Mandelstam-Leibbrandt prescription in
higher-dimensional perturbative gauge theory.

\end{abstract}
\pacs{PACS numbers: 11.10.Kk, 11.10.St, 11.15.Pg, 11.15.-q, 12.38.Cy
\vspace{1.5cm}\\
{\small \hspace*{-1.7cm}CERN-TH/97-245\\
\hspace*{-1.7cm}September 1997}}
\narrowtext

QCD$_2$, the dynamical theory of quarks and gluons in
two dimensions, has been a fascinating testing-ground for some of the
important concepts of high energy physics.
A profound analysis of the theory was given by 't Hooft \cite{1}
in the  limit
$N \rightarrow \infty$ corresponding to an infinite number of colors.
't Hooft's study has culminated in the explicit demonstration of
quark confinement and in the numerical determination of the bound state
meson spectrum.

Technically, 't Hooft has worked in a generalized axial gauge, which
means that the gauge potential is set to zero along a fixed
direction. This direction is most conveniently chosen as light-like.
Within this so-called light-cone gauge
($A_{-}=0$), one of the two quark spinor components decouples.
't Hooft has treated the remaining infra-red singularity of the gauge
field propagator in a way equivalent (cf \cite{2}) to
a principal part prescription:
\begin{equation}
\label{thooft}
\frac{1}{k^2_-} \rightarrow
\frac{{\cal P}}{k^2_-}=\frac{1}{2} \Bigg[ \frac{1}{(k_- +i \epsilon)^2} +
\frac{1}{(k_- -i \epsilon)^2} \Bigg].
\end{equation}

The highly consistent 't Hooft theory has been extended in many
studies. In particular,  the invariance of the meson spectrum with
respect to a different choice of gauge, the space-like axial gauge,
has been checked
explicitly by a combination of analytical and
numerical calculations\cite{3}.  In this more
complicated gauge one has to deal with both quark spinor components,
and the integral equation for the meson bound states becomes
two-dimensional. Nevertheless, an identical bound state meson spectrum
was obtained \cite{3}.

QCD$_2$ would appear to be completely satisfactory were it not for
two potential
problems, which turn out to be closely related. The first is, that
{\it using the principal part prescription~(\ref{thooft}), Wick rotation
to Euclidean space is impossible} \cite{remark}.
This was remarked early on by
Wu \cite{wu}, who suggested to formulate QCD$_2$ in Euclidean space.
He considered the Wick-rotated kernel
($k_-=k_1-k_0 \rightarrow k_1+i k_2$)
\begin{equation}
\label{wu}
\frac{1}{k^2_-} \rightarrow \frac{1}{(k_1+i k_2)^2},
\end{equation}
and proposed to use symmetric integration. This
means that the integrals with the kernel eq.~(\ref{wu}) should
exclude a small spherical hole of radius $\epsilon$ around the
origin of the $(k_1,k_2)$ plane.

Secondly, {\it the principal part
prescription for generalized axial gauges appears to be irreparably
inconsistent above two dimensions} since
manifest perturbative renormalizability is lost.
For light-cone gauges above two dimension, the regularization
suggested by Mandelstam \cite{8} and by Leibbrandt \cite{9}
appears to solve this problem
({\em cf}  also \cite{capper}).
In two dimensions, this prescription gives
\begin{equation}
\label{ml}
\frac{1}{k^2_-} \rightarrow \frac{1}{(k_- - i \epsilon~{\rm sgn} k_+
)^2}=
\frac{{\cal P}}{k^2_-} - i \pi~\delta'(k_-)~{\rm sgn} k_+.
\end{equation}
Bassetto et al.~\cite{bassetto}  noted that the Wu
kernel (\ref{wu}) and the Mandelstam-Leibbrandt kernel (\ref{ml}) are
related to each other by a Wick rotation.

Wu \cite{wu} derived an effective integral equation
for the bound state mesons
in QCD$_2$ at $N=\infty$ with the regularization of
eqs.~(\ref{wu}), (\ref{ml}).
Physical observables  should not depend on the regularization
scheme, and it was widely suspected that the Wu-Mandelstam-Leibbrandt
(WML) kernels should lead to the same results
as the 't Hooft prescription.  Past efforts to solve the Wu equation
have however failed.

The apparent impossibility to resolve the meson bound state equation
in the Wu formalism is not a true impediment to study QCD$_2$ in the
WML prescription.
Complementary insight can be obtained
by probing the Yang-Mills field by static flux lines instead
of dynamical quarks: the Wilson loops. This has already been the basic
motivation of work by Bassetto et al.~\cite{bassetto}.

The present paper consists in a study of Wilson loops for QCD$_2$.
We first investigate perturbatively various contours in
Euclidean and Minkowski space, up to ${\cal O}(g^6)$,
using numerical and analytic calculations. We then
derive the exact non-perturbative result.
Our study confirms explicitly the finding \cite{bassetto}
that the WML regularization
gives a result {\em different} from 't~Hooft's.  However,
contrary to what was claimed before in the literature \cite{bassetto},
the result fulfills the same consistency conditions as 't~Hooft's
Wilson loop: {\em (i)} it does not depend on the
shape of the closed contour, but only on its area,
{\em (ii)} the leading behavior at finite $N$ is an area law behavior,
indicating confinement, and {\em (iii)} the result is gauge invariant
in a (restricted) way which will be detailed later on.
In our opinion, these tests put the  WML regularization on an equal
footing with the 't Hooft prescription.

Before endeavoring on the detailed calculations in the WML
regularization, we briefly present the precise definitions and
review the main results of the analogous calculation using 't
Hooft's prescription.
Wilson loops are defined as
\begin{equation}
\label{wilson}
{\cal W}_{{\cal C}}=
\langle \frac{1}{N} P {\rm Tr} \exp \big( i g \oint_{{\cal C}} d\vec{x}
\cdot \vec{A} \big)
\rangle.
\end{equation}
Here $P$ denotes path ordering of the gauge field $\vec{A}$ along the
contour ${\cal C}$.  In two Euclidean dimensions the standard,
exact result can
be  obtained in a gauge invariant way for any number of
colors and by a variety of
methods; {\em e.g.} using a manifest gauge invariant formulation like
lattice gauge theory.
For the case of a simple
(i.~e.~not self-intersecting) $U(N)$ or $SU(N)$ Wilson loop
${\cal W}_{{\cal C}}$ along a contour ${\cal C}$ one obtains
\begin{equation}
\label{expo}
{\cal W}_{{\cal C}}^{U(N)}=e^{- \frac{1}{2} N g^2 {\cal A}_{{\cal
C}}},
\;\;
{\cal W}_{{\cal C}}^{SU(N)}=e^{- \frac{1}{2} (N-\frac{1}{N}) g^2
{\cal A}_{{\cal C}}},
\end{equation}
where ${\cal A}_{{\cal C}}$ is the area enclosed by ${\cal C}$.
Apart from the trivial
factor $N$, the $U(N)$ result does not distinguish between the
Abelian and the non-Abelian case. Furthermore,
the form of eq.~(\ref{expo}) agrees with the
idea of a linear confining potential between sources.
${\cal W}_{{\cal C}}$ depends on the contour ${\cal C}$ solely through
the dimensionless combination $g^2 {\cal A }_{{\cal C}}$.
The general reasons for this fact
were first emphasized in \cite{witten}:
YM$_2$ is invariant under
area preserving diffeomorphisms \cite{foot}. Even in Minkowski space,
the 't Hooft prescription (\ref{thooft}) (or
any other generalized axial gauge with principal part prescription)
reproduces the exponential law (\ref{expo}) with
${\cal A }_{{\cal C}} \rightarrow i \tilde{{\cal A }}_{{\cal C}}$ where
$\tilde{{\cal A }}_{{\cal C}}$ is the Minkowski ``area'' enclosed by
the contour.

Let us first work in Euclidean space and use the prescription (\ref{wu}).
The path-ordered exponential in eq.~(\ref{wilson}) is defined as
\begin{equation}
\label{path}
{\cal W}_{{\cal C}}=\frac{1}{N}
\sum_{n=0}^{\infty} (-g^2)^n \int_0^1 ds_1
\dot{x}_-(s_1) \ldots
\int_0^{s_{2 n-1}} ds_{2 n} \dot{x}_-(s_{2 n})
{\rm Tr} \langle A_+\big(\vec{x}(s_1)\big) \ldots
A_+\big(\vec{x}(s_{2 n})\big)
\rangle,
\end{equation}
where $\vec{x}(s)$, $s \in [0,1]$ parametrizes the closed contour
${\cal C}$. The expectation value in eq.~(\ref{path}) is to be evaluated
by the Wick rule, and the basic correlator, obtained from
eq.~(\ref{wu}) by simple Fourier transform to configuration space, is
$(x_{\pm}=x_1 \mp i x_2)$
\begin{equation}
\label{wick}
\langle (A_+)_{i,j}(\vec{x}) (A_+)_{k,l}(\vec{x'}) \rangle =
\frac{1}{4 \pi}~\delta_{i,l} \delta_{j,k}~\frac{x_+ - x'_+}{x_- - x'_-}.
\end{equation}
Here we have also written out the matrix indices of the gauge field
$A_+$.

It is simplest to consider a circular contour.
Then the weighted basic correlator is independent of the variables
$s,s'$:
$\dot{x}_-(s)~\dot{x}_-(s')~(x_+(s)-x_+(s'))/(x_-(s)-x_-(s'))=(2 \pi
r)^2$.
Now the integration over the path parameters $s_1,\ldots,s_{2 n}$
becomes trivial, and the computation can be very simply performed,
say up to three loops. One finds for $U(N)$
\begin{equation}
\label{threeloop}
{\cal W}_{{\cal C}}^{U(N)}=
1-\frac{1}{2}~N g^2 {\cal A}_{{\cal C}}
+\frac{1}{8} (\frac{2}{3} N^2+ \frac{1}{3})~g^4 {\cal A}_{{\cal C}}^2
-\frac{1}{48} (\frac{5}{15} N^3
+ \frac{10}{15} N)~g^6 {\cal A}_{{\cal C}}^3 + {\cal O}(g^8),
\end{equation}
where ${\cal A}_{{\cal C}}=\pi r^2$ is the area of the circle.
Evidently this three-loop result is incompatible with the standard
law eq.~(\ref{expo}) unless $N=1$. The technical reason is that the
Abelian result is reproduced in the WML prescription
by non-planar gluon exchange; as soon as $N>1$, crossing gluon lines
acquire a special weight, as is evident from eq.~(\ref{threeloop}):
At ${\cal O}(g^4)$, {\em e. g.},  there are two planar and one crossed
diagram, whereas at ${\cal O}(g^6)$ five planar and ten crossed
diagrams are present.

Is the result eq.~(\ref{threeloop}), derived for a circular contour,
generally true for all simple loops of the same area?
Does eq.~(\ref{threeloop}) remain valid for contours in
Minkowski space, if we perform the analytic continuation
${\cal A }_{{\cal C}} \rightarrow i \tilde{{\cal A }}_{{\cal C}}$?
The answer to both questions is {\it yes}.
The argument for the first affirmation is the invariance under area preserving
diffeomorphisms \cite{foot}, which should not depend on the
regularization scheme. The second statement should be true since
the kernels (\ref{wu}) and (\ref{ml}) are related by
analytic continuation, and since
the Wu kernel (\ref{wu}) possesses a commutativity property
already emphasized in the original work \cite{wu}.

Since the two above assertions are crucial for the following,
we have convinced
ourselves of their validity
for a variety of simple contours such as ellipses,
triangles, rectangles of various orientations both in Euclidean and
in Minkowski space. Careful numerical evaluation of eq.~(\ref{path}) up
to ${\cal O}(g^4)$ and in some case up to
${\cal O}(g^6)$ yielded in all cases perfect
agreement with eq.~(\ref{threeloop}) to a level of precision
of $10^{-4} \ldots 10^{-3}$. Our numerical computations thus lend
strong support to the formal general arguments mentioned above.

Our findings are seemingly in disagreement with the analytical
two-loop (i.e.~${\cal O}(g^4)$) calculations of Bassetto
et al.~\cite{bassetto} for rectangular contours in Minkowski space
\cite{notation}. They consider two orientations of the loop.
In the first case, the rectangle is oriented along the light-cone
space and light-cone time axes.
Agreement with eq.~(\ref{threeloop}) can be established
\cite{acknowledge}.
In the second case, the rectangle is oriented along the space and
time axes. Here they concluded that the WML regularization violated
the area law. However, the
authors failed to realize that the complicated dependence of their
final result on the aspect ratio of their rectangle exactly {\it
cancels}, as is easily verified. If this cancellation is taken into
account, the Bassetto et~al.~result is
in perfect agreement with ours.

We also investigated the invariance of ${\cal W}_{{\cal C}}$ under
transformations to generalized axial gauges (cf \cite{wu})
\begin{equation}
\frac{1}{(k_1+i k_2)^2} \rightarrow \frac{1}{(k_1 \cos \theta +i
k_2 \sin \theta)^2}
\end{equation}
for $0<\theta\leq \pi/4$. In that case, symmetric integration
amounts to cutting out a spherical hole in the
$(\tilde{k}_1,\tilde{k}_2)$ plane with $\tilde{k}_1= k_1 \cos \theta$ and
$\tilde{k}_2= k_2 \sin \theta$. The propagator becomes
$x_+/x_- \rightarrow \tilde{x}_+/\tilde{x}_-$,
with $\tilde{x}_{\pm} = x_1/\cos \theta \mp i
x_2/\sin \theta$. As before, the gauge invariance was
checked numerically to a relative precision of
about $10^{-3}$ for several values of $\theta$.

Having established that the choice of contour is arbitrary,
we may restrict ourselves to the especially simple Euclidean circular
contour. There we can find all further terms in eq.~(\ref{threeloop}),
since the integrand in eq.~(\ref{path}) is constant.
The problem of determining the Wilson
loop reduces to the purely combinatorial problem of finding the
group-theoretic factors corresponding to the Wick contractions.
Fortunately, these factors are generated by a simple matrix integral:
\begin{equation}
\label{matrix}
{\cal W}_{{\cal C}}=\frac{1}{Z} \int {\cal D} F~\exp \big(- \frac{1}{2}
{\rm Tr} F^2 \big)~\frac{1}{N}
{\rm Tr} \exp \big(i g \sqrt{{\cal A }_{{\cal C}}} F \big).
\end{equation}
For $U(N)$, ${\cal D} F$ denotes the flat integration measure on the space of
hermitian $N~\times~N$ matrices:
${\cal D} F=\prod_{i=1}^{N} dF_{ii} \prod_{i<j}^{N} d({\rm Re}F_{ij})
d({\rm Im}F_{ij})$. $Z$ is a normalization factor:
$Z=\int {\cal D} F~\exp \big(- \frac{1}{2} {\rm Tr} F^2 \big)$.
This matrix integral has been evaluated
with a variety of methods \cite{brezin}. The
final result constitutes the exact expression for the Wilson
loop at any $N$:
\begin{equation}
\label{contour}
{\cal W}_{{\cal C}}^{U(N)}=\exp \big(- \frac{1}{2} g^2 {\cal A }_{{\cal
C}}\big)~\frac{1}{N}
\oint \frac{dz}{2 \pi i}~\exp \big( -g^2 {\cal A }_{{\cal C}} z \big)
\bigg( \frac{z+1}{z} \bigg)^N.
\end{equation}
The contour integral, which encloses the multiple pole at $z=0$,
gives a Laguerre polynomial in $g^2 {\cal A }_{{\cal C}}$ of order
$N-1$:
$ L^1_{N-1}(g^2 {\cal A }_{{\cal C}})$.
The first few examples are:
\begin{equation}
\label{examples}
{\cal W}_{{\cal C}}^{U(1)}=e^{- \frac{1}{2} g^2 {\cal A }_{{\cal
C}}},\;\;
{\cal W}_{{\cal C}}^{U(2)}=\Big(1-\frac{1}{2}
g^2 {\cal A }_{{\cal C}}\Big)e^{- \frac{1}{2}
g^2 {\cal A }_{{\cal C}}},\;\;
{\cal W}_{{\cal C}}^{U(3)}=\Big(1- g^2 {\cal A }_{{\cal C}}
+\frac{1}{6}(g^2 {\cal A
}_{{\cal C}})^2\Big)
e^{- \frac{1}{2} g^2 {\cal A }_{{\cal C}}}.
\end{equation}
For $SU(N)$, the integration measure in eq.~(\ref{matrix}) has to be
modified to enforce tracelessness:
${\cal D} F \rightarrow {\cal D} F~\delta({\rm Tr} F)$.
The $U(1)$ part decouples from the $U(N)$ Wilson loop and we find
\begin{equation}
\label{sun-result}
{\cal W}_{{\cal C}}^{SU(N)}=\exp \big( \frac{1}{2 N} g^2 {\cal A
}_{{\cal C}} \big)~
{\cal W}_{{\cal C}}^{U(N)}.
\end{equation}
Our result coincides with the usual expression (\ref{expo}) only
in the Abelian $U(1)$ case. Incidentally,
this explains why Wu and Stamatescu \cite{12}
were able to reproduce  the standard solution of the Schwinger model
(i.e.~two-dimensional QED) with the WML regularization.

We now study the exact formulas (\ref{contour}),
(\ref{sun-result})
in 't Hooft's large $N$ limit: $N \rightarrow \infty$, $g \rightarrow
0$,
with fixed $\tilde{g}^2 = N g^2$.
Eqs.~(\ref{contour}), (\ref{sun-result}) become the integral representation
of a Bessel function:
\begin{equation}
\label{Nresult}
{\cal W}_{{\cal C}}^{U(\infty)} =
{\cal W}_{{\cal C}}^{SU(\infty)} =
\frac{1}{\tilde{g} \sqrt{{\cal A}_{{\cal C}}}}~
J_1(2 \tilde{g} \sqrt{{\cal A}_{{\cal C}}}).
\end{equation}
Curiously, the factors
$\exp (-\frac{1}{2} g^2 {\cal A}_{{\cal C}})$
present at all finite $N$ (cf eqs.~(\ref{contour}), (\ref{examples}))
disappear in the $N \rightarrow \infty$ limit. But it is precisely these
factors which lead to confinement at finite $N$. Very much unlike the
usual theory, where from eq.~(\ref{expo}) we have
${\cal W}_{{\cal C}}^{U(\infty)}=
\exp(-\frac{1}{2} \tilde{g}^2 {\cal A}_{{\cal C}})$,
in the present model
the behavior changes qualitatively at $N=\infty$, as
${\cal A }_{{\cal C}} \rightarrow \infty$
\begin{equation}
\label{asymp}
{\cal W}_{{\cal C}}^{U(\infty)}
\sim \frac{1}{\sqrt{\pi}} (\tilde{g}^2 {\cal A }_{{\cal
C}})^{-\frac{3}{4}}
\cos (2 \tilde{g} \sqrt{{\cal A }_{{\cal C}}} - \frac{3}{4} \pi)
\end{equation}
This fall-off at large areas ${\cal A }_{{\cal C}}$ is
far too slow to ensure confinement! We therefore conclude that Wu's
model of mesons {\em has to be studied at finite $N$}; the Bethe-Salpeter
equation written down in \cite{wu} is not expected to lead to a
discrete meson spectrum.

What have we achieved so far? We have presented evidence that, besides
the 't Hooft principal value prescription, the WML symmetric integration
yields a consistent yet different theory. We have found this theory
to be qualitatively different in the $N\rightarrow \infty$ limit.
This allows us to solve the old riddle of why bound states for the Wu
equation have never been found. It also allows us to expose a
case in which the $N\rightarrow \infty$ limit is {\em invalid}.

At finite $N$, we are lead to conclude that two physically different,
consistent, gauge theory formulations in
$D=2$ exist, unless inconsistencies of the WML
prescription end up being found
on a subtler level. It is indeed known that QCD$_2$ is special
in that it can be considerably generalized by adding higher powers
(products of traces) of the field strength to the action; such terms
are irrelevant in $D>2$ but begin to scale at exactly $D=2$
\cite{witten},\cite{gQCD}. Therefore, as already pointed out in
\cite{gQCD}, continuing a $D$-dimensional formulation of QCD
down to $D=2$ might not give the usual Tr$F^2$ theory, but
one of the generalized theories (gQCD$_2$) \cite{conjecture}.
However, this does not explain why there are coexisting
theories with the {\it same} Lagrangian
${\cal L} = -\frac{1}{4} {\rm Tr} F^2$. It is easily seen that
the WML prescription in Euclidean space cannot be described
by a standard (even generalized) lattice gauge theory.
It would be very interesting to find a lattice discretization
for this theory.

We emphasize that the two prescriptions naturally result from different
points of view: the standard prescription is obtained from
a number of strictly two-dimensional quantization methods
(Euclidean lattice gauge theory, light-cone quantization)
while the WML prescription seems compelling from
the viewpoint of higher dimensional Minkowski perturbative QCD.
It appears that ``theory space'' is big enough in $D=2$ to
allow coexistence of these theories.

\acknowledgements
We would like to thank I.~Bars, A.~Bassetto, S.~Wadia and especially
T.~T.~Wu for several inspiring discussions.
W. K. thanks the CERN theory division for hospitality.


\begin{references}
\bibitem{remark}
      This problem does not seem to arise in the spatial axial
      gauge treatment of ref. \cite{3}. However, in order
      to achieve agreement with the light-cone gauge result,
      an unusual $i \epsilon$ prescription for the
      quark propagators was necessary. This again interferes with
      Wick rotation.
\bibitem{foot}
      This can be seen by realizing that
      the field strength is a {\em two-form} in two dimensions.
      The action is therefore invariant under all diffeomorphisms
      that preserve the volume element of the surface.
\bibitem{notation}
      In order to compare, one has to make the trivial change
      $U(N) \rightarrow SU(N)$ (see eq.~(\ref{sun-result})).
      An additional replacement  $g^2 \rightarrow \frac{1}{2} g^2$
      is due to a difference in normalization
      of the group's generators.
\bibitem{conjecture} We conjecture that these generalizations are
      responsible for further disagreements \cite{bassetto}
      between $2+\epsilon$
      calculations in Feynman gauge and calculations in strictly
      $D=2$.
\end{references}
\end{document}